\RequirePackage{fix-cm}
\documentclass[smallextended]{svjour3}       
\smartqed  
\usepackage{graphicx}
\begin{document}

\title{New Superpotential in Conservation Laws in General Relativity
}

\author{Jiri Adamek         \and
        Jan Novotny 
}


\institute{Institute of Plasma Physics of the CAS \at
              Za Slovankou 3, Prague, Czech Republic \\
              Tel.: +420-6605-2961\\
              \email{adamek@ipp.cas.cz}           
           \and
           Faculty of Science, Masaryk University \at
           Kotlarska 2, Brno, Czech Republic
}

\date{Received: date / Accepted: date}

\maketitle

\begin{abstract}
This work refers to the new formula for the superpotential $U_i^{kl}$ in conservation laws in general relativity satisfying the integral and differential conservation laws within the Schwarzschild metric. The new su-perpotential is composed of two terms $U_i^{kl} =M_i^{kl} + A_i^{kl}$. The $M_i^{kl}$ is based on M$\o$ller's concept [C. M$\o$ller, Ann. Phys. (NY) 4, 347 (1958)] and it is a function of the metric $g_{ik}$ and its first derivative only. The second term $A_i^{kl}$ is the antisymmetric tensor density of weight +1 and it consists of higher derivatives of the metric $g_{ik}$ . Although the new superpotential consists of higher derivatives of the metric $g_{ik}$ it might bring a new evaluation of the conservative quantities in general relativity.\keywords{superpotential \and conservation laws \and energy-momentum complex}
\end{abstract}

\section{Introduction}
\label{intro}
The recent  observation of gravitational waves \cite{Abbott} might increase further investigation of energy-momentum complex $\tau_i^k$ within the conservation laws in general relativity \cite{Horsky,Chandrasekhar}, which is based on the definition of the superpotential $U_i^{kl}$ where $\tau_i^k = U^{kl}_{i},_l$. The variety of energy-momentum complex $\tau_i^k$ based on a different concept like Landau-Lifshits \cite{Landau,Virbhadra1}, Einstein-Freud \cite{Moller1}, M$\o$ller \cite{Moller1,Novotny,Moller2}, Mickevic \cite{Moller2} can be used to study the conservative quantities \cite{Favata,Virbhadra,Nashed} or gravitational waves \cite{Su}. Therefore, we can also find a different formula of the superpotential $U_i^{kl}$ \cite{Stolin,Novotny2,Komar,Francavigli}. The gravitational energy-momentum problem has been also discussed within the teleparallel gravity \cite{Andrade}. However, a different construction of the superpotential $U_i^{kl}$ causes a problem in conservation laws in general relativity and it raises the question what is the ``right'' superpotential? The energy-momentum complex must yield the total energy of the whole system equal to $Mc^2$ and it must be invariant with respect to the Lorentz transformation and independent on the local coordinate system. This defines the integral and differential conservation laws in general relativity. It is shown that none of the above-mentioned candidates of energy momentum complex can satisfy all conservation laws. Therefore, we introduce the novel concept of the superpotential, which is composed of a two parts $U_i^{kl} =M_i^{kl} + A_i^{kl}$. The first term $M_i^{kl}$, is based on the M$\o$ller's concept and it is a function of the metric $g_{ik}$ and its first derivative only. The second term $A_i^{kl}$ is more complex and it consists of higher derivatives of the metric. It must be noted, that the formula for $A_i^{kl}$ was found using a deductive and intuitive method. However, the novel concept of a superpotential satisfying the integral and differential conservations laws can contribute to the understanding of the gravitational energy in general relativity. A standard classification of the known superpotentials is based on the formula of the general superpotential \cite{Novotny2}

\begin{eqnarray}
\label{generalsup}
U_i^{kl} &=& \frac{1}{2\kappa}\frac{c_1}{\alpha}\left(-g\right)^{\alpha},_m\left(g^{lm}\delta^{k}_{i}-g^{km}\delta^{l}_{i}\right)+
\frac{c_2}{2\kappa}\left(-g\right)^{\alpha}\left(g^{lm},_m\delta^{k}_{i}-g^{km},_m\delta^{l}_{i}\right)+ \nonumber \\ 
&& \frac{c_3}{2\kappa}\left(-g\right)^{\alpha}g_{ib},_m\left(g^{bl}g^{mk} - g^{bk}g^{ml}\right)
\end{eqnarray}

with constant $\kappa=8\pi Gc^{-4}$ and the gravitational constant $G$. The general superpotential is composed of a linear combination of three terms, which are functions of the metric $g_{ik}$ and its first derivative. The $U_i^{kl}$ consists of four constants $c_1$, $c_2$, $c_3$ and $\alpha$, which can be chosen in order to define a different energy-momentum complex $\tau_i^k$. The following table represents the values of all constants mentioned above with respect to a different energy-momentum complex.
\begin{table}
\caption{Variety of the energy momentum complex and its constants in the formula for the general superpotential}
\label{tab:1}       
\begin{tabular}{ccccc}
\hline\noalign{\smallskip}
  & Einstein-Freud & Landau-Lifshitz & M$\o$ller & Mickevic   \\
\noalign{\smallskip}\hline\noalign{\smallskip}
$c_1$ & 1 & 1 & 0 & 0  \\
$c_2$ & 1 & 1 & 0 & 0  \\
$c_3$ & 1 & 1 & 2 & 1  \\
$\alpha$ & 0.5 & 1 & 0.5 & 0.5  \\
\noalign{\smallskip}\hline
\end{tabular}
\end{table}

\section{Application of the integral and differential conservation laws in Schwarzschild metric}
\label{conservative_laws}

The general superpotential $U_i^{kl}$ (\ref{generalsup}) is assumed to be the most proper candidate for the evaluation of the energy-momentum complex $\tau_i^k$ or the general relativity conservation integral quantity four-momentum $P_i$ 
\begin{equation}
\label{Pi}
P_i=\frac{1}{c}\mathop{\oint}_{\partial V} U_i^{0\lambda} \mathrm{d}\sigma_{0\lambda}
\end{equation}
The four unknown constants can be determined by using differential and integral conditions. There exist two integral conditions in conservation laws. The first one demands the total energy of the insular system ($\partial V\rightarrow \infty$) to be $E=Mc^2$ or $P_0=Mc$, which can be expressed as

\begin{equation}
\label{Int1}
P_i=\frac{1}{c}\mathop{\oint}_{\partial V\rightarrow \infty}U_i^{0\lambda} \mathrm{d}\sigma_{0\lambda}=Mc\delta^{0}_{i}
\end{equation}
$$\lambda =1,2,3\ \ \ i=0,1,2,3 $$
The second integral conservation law requires the four-momentum be invariant with respect to the Lorentz transformation. This leads to the condition
\begin{equation}
\label{Int2}
\frac{1}{c}\mathop{\oint}_{\partial V\rightarrow \infty}U_i^{k\lambda} \mathrm{d}\sigma_{0\lambda}=0
\end{equation}
$$\lambda , \kappa =1,2,3\ \ \ i=0,1,2,3 $$
If we assume a metric $g_{ik}$ of an external Schwarzschild solution \cite{Inverno} with stationary gravitation field in spherical coordinating system $x^i=\left(ct, r, \varphi,    \vartheta \right)$
\begin{equation}
\label{Schwar}
g_{00}=\left(1-\frac{2m}{r}\right),\; g_{11}=-\left(1-\frac{2m}{r}\right)^{-1},\; g_{22}=-r^2sin^{2}\vartheta, \; g_{33}=-r^2  
\end{equation}
with $m=G\cdot M/c^2$ then both integral conservation laws, (\ref{Int1}) and (\ref{Int2}), using the general superpotential (\ref{generalsup}) can be expressed by two linear equations  

\begin{equation}
\label{lin1}
c_1-\frac{c_2}{2}+\frac{c_3}{2}=1
\end{equation}
\begin{equation}
\label{lin2}
2c_1-c_2-c_3=0
\end{equation}
with three unknown constants $c_1$, $c_2$, $c_3$. The remaining coefficient $\alpha$ in (\ref{generalsup}) is a free parameter and its value is chosen to be 0.5. Finally, the third equation will be found by using a differential condition, which states that the evaluation of the four-momentum $P_i$ must be independent of the spatial coordinate. Only the third term of the superpotential (\ref{generalsup}) is independent of the spatial coordinate. Then the values of constants $c_1$, $c_2$ must be equal to zero in order to satisfy the differential conservation law. 
Finally, we obtain from (\ref{lin1}) $c_3=2$ and from (\ref{lin2}) $c_3=0$. In this case, the system of our equations is insolvable and therefore equation (\ref{generalsup}) does not represent the proper formula of the superpotential. This is the main problem of conservation laws in general relativity.

\section{The novel concept of superpotential }
\label{newsuperpotential}

A novel form of the superpotential has been found under the assumption that also higher derivatives of the metric $g_{ik}$ should be taken into account. We propose here the new superpotential
\begin{equation}
\label{newsup}
U_i^{kl} =M_i^{kl} + A_i^{kl}
\end{equation}
where
\begin{equation}
\label{newsup_M}
M_i^{kl}=\frac{c_{3}}{2\kappa}\sqrt{-g}\:g_{ib},_m\left(g^{bl}g^{mk} - g^{bk}g^{ml}\right)
\end{equation}
\begin{equation}
\label{newsup_A}
A_i^{kl}=\frac{c_4}{2\kappa}\sqrt{-g}\:C\:\!S,_m\left(g^{lm}\delta^{k}_{i}-g^{km}\delta^{l}_{i} \right)
\end{equation}
\begin{equation}
\label{C}
C=\left(S\right)^{\alpha} \left(S,_a\!S,_b\!g^{ab}  \right)^{\beta},\ \ \ \alpha=\frac{3}{2}, \ \beta=-1
\end{equation}
\begin{equation}
\label{S}
S=R^{abcd}R_{abcd}
\end{equation}
The new concept of the superpotential is based on a linear combination of two terms. The first term $M_i^{kl}$(\ref{newsup_M}) is based on M$\o$ller's concept and it is a function of the metric $g_{ik}$ and its first derivative. It corresponds to the third term in the general formula of the superpotential (\ref{generalsup}). The second term $A_i^{kl}$(\ref{newsup_A}) is an antisymmetric tensor density of weight +1. It is composed of scalar quantity $C$ (\ref{C}) and the Reimann tensor scalar invariant $S$ (\ref{S}) \cite{Inverno}. Therefore, the calculation of the four-momentum $P_i$ (\ref{Pi}) will be independent of the coordinate system, which satisfies the differential condition. This is the main idea of adding term $A_i^{kl}$. The final formula of the new superpotential (\ref{newsup}) was found by a rather intuitive way and it consists of a higher derivative of the metric $g_{ik}$. This would require a Langrangian of the gravitational field $\sqrt{-g}L\left(g_{ik},_{lm};g_{ik},_{l};g_{ik}\right)$ \cite{Inverno} with higher derivatives. Despite the previous argument being non-trivial \cite{Babak}, we would then be able to satisfy both integral conservation laws (\ref{Int1}) and (\ref{Int2}) by searching for constant $c_3$ and $c_4$ in (\ref{newsup_M}) and (\ref{newsup_A}), respectively. The scalars $C$ and $S$ \cite{Inverno} in spherical coordinates in the Schwarzschild metric $g_{ik}$ (\ref{Schwar}), assuming the asymptotic approximation $r\rightarrow \infty$ giving
\begin{equation}
\label{SC}
S =48\frac{m^2}{r^6};\ C=-\frac{1}{36\sqrt{48}}\frac{r^5}{m}
\end{equation}
Then the first integral conservation law (\ref{Int1}) can be expressed as a linear equation
\begin{equation}
\label{lin3}
\frac{c_3}{2}-\frac{c_4}{2\sqrt{12}}=1
\end{equation}
The second integral conservation law (\ref{Int2}) is more complex and can be expressed \cite{Moller2} as
\begin{equation}
\label{Int2_new}
\mathop{\oint}_{\partial V\rightarrow \infty}M_i^{k\lambda} \mathrm{d}\sigma_{0\lambda}+\mathop{\oint}_{\partial V\rightarrow \infty}A_i^{k\lambda} \mathrm{d}\sigma_{0\lambda}
=-c_3\frac{1}{3}Mc^{2}\delta^{k}_{i}-c_4\frac{1}{3\sqrt{12}}Mc^{2}\delta^{k}_{i}=0,
\end{equation}
which yields the second linear equation 
\begin{equation}
\label{lin4}
c_3+\frac{c_4}{\sqrt{12}}=0
\end{equation}
The linear equations (\ref{lin3}) and (\ref{lin4}) provide the constants   
\begin{equation}
\label{lin5}
c_3=1\ \ \ c_4=-\sqrt{12}
\end{equation}
It is also interesting that the total energy of the insular system ($\partial V\rightarrow \infty$) as calculated in (\ref{lin3}) is divided in two identical parts described by $M_i^{kl}$ and $A_i^{kl}$. The final formula of the new superpotential is
\begin{equation}
\label{newsup_formula}
U_i^{kl}=\frac{\sqrt{-g}}{2\kappa}\left[g_{ib},_m\left(g^{bl}g^{mk} - g^{bk}g^{ml}\right)+ \sqrt{12}\:C\:\!S,_m\left(g^{km}\delta^{l}_{i}-g^{lm}\delta^{k}_{i}\right)\right]
\end{equation}

\section{Conclusions}
\label{Conclusion}
It was shown that none of the linear combinations of the three terms in the general superpotential can yield a superpotential which satisfies all differential and integral conditions in the case of the Schwarzschild metric. In this paper we introduce a novel concept of the superpotential, which is based on the linear combination of M$\o$ller's concept $M_i^{kl}$, consisting only of the first derivative of the metric $g_{ik}$, and an additional term $A_i^{kl}$ with a higher derivative of the metric $g_{ik}$. It seems that the presence of higher derivatives of the metric in the superpotential is necessary in order to satisfy the differential and both integral conditions. This would require a Langrangian of the gravitational field, which includes derivatives of the metric $g_{ik}$ higher than second order.   

\begin{acknowledgements}
I would like thank to Dr. Jana Jurmanova and Dr. Jan St\"{o}ckel for the fruitful discussion and to Dr. Vladimir Fuchs for correcting English and to my family supporting me in my work.
\end{acknowledgements}

\end{document}